\documentclass[fleqn,usenatbib]{mnras}

\usepackage{newtxtext,newtxmath,comment}

\usepackage[T1]{fontenc}

\DeclareRobustCommand{\VAN}[3]{#2}
\let\VANthebibliography\thebibliography
\def\thebibliography{\DeclareRobustCommand{\VAN}[3]{##3}\VANthebibliography}

\usepackage{graphicx}	
\usepackage{amsmath}	

\def\xmm {\emph{XMM--Newton}}
\def\nustar {\emph{NuSTAR}}

\title[X-ray/IR correlated variability in 4U 1728-34]{Sub-second infrared variability from the archetypal accreting neutron star  4U~1728-34}
\author[Vincentelli et al.]{F.M.  Vincentelli$^{1,2}$, P. Casella$^{3}$, A. Borghese$^{1,2}$, Y. Cavecchi$^{4}$, G. Mastroserio$^{5}$, L. Stella$^{3}$, D. Altamirano$^{6}$, \newauthor M. Armas Padilla$^{1,2}$, M. C. Baglio$^{7,8}$, T. M. Belloni$^{8}$, J. Casares$^{1,2}$, V. A. C\'uneo$^{1,2}$, N. Degenaar$^{9}$, \newauthor M. D{\'i}az Trigo$^{10}$, R. Fender$^{11}$, T. Maccarone$^{12}$, J. Malzac$^{13}$, D. Mata S\'anchez$^{1,2}$, M. Middleton$^{6}$, S. Migliari$^{14,15}$,  \newauthor  T. Muñoz-Darias$^{1,2}$, K. O’Brien$^{16}$, G. Panizo-Espinar$^{1,2}$, J. Sánchez-Sierras$^{1,2}$, D. M. Russell$^{7}$, P. Uttley$^{9}$
\\
$^{1}$Instituto de Astrof\'{i}sica de Canarias, E-38205 La Laguna, Tenerife, Spain\\
$^{2}$Departamento de Astrof\'{ı}sica, Universidad de La Laguna, E-38206 La Laguna, Tenerife, Spain\\
$^{3}$INAF, Osservatorio Astronomico di Roma 
Via Frascati 33, I-00078 Monteporzio Catone, Italy\\
$^4$Departament de Física, EEBE, Universitat Politècnica de Catalunya, Barcelona, Spain\\
$^5$INAF – Osservatorio Astronomico di Cagliari, via della Scienza 5, I-09047 Selargius (CA), Italy\\
$^{6}$Department of Physics and Astronomy, University of Southampton, SO17 1BJ, UK \\
$^7$Center for Astro, Particle and Planetary Physics, New York University Abu Dhabi, PO Box 129188, Abu Dhabi, UAE\\
$^8$INAF - Osservatorio Astronomico di Brera, Via E. Bianchi 46, I-23807 Merate, Italy \\
$^9$Anton Pannekoek Institute for Astronomy, University of Amsterdam, Science Park 904, 1098 XH Amsterdam, The Netherlands \\
$^{10}$ESO, Karl-Schwarzschild-Strasse 2, 85748 Garching bei M\"unchen, Germany\\
$^{11}$Astrophysics, Department of Physics, University of Oxford, Denys Wilkinson Building, Keble Road, Oxford OX1 3RH, UK \\
$^{12}$Department of Physics and Astronomy, Texas Tech University, Lubbock, TX 79409-1051, USA \\
$^{13}$IRAP, Universite' de Toulouse, CNRS, UPS, CNES, Toulouse, France\\
$^{14}$XMM-Newton Science Operation Center, ESAC/ESA, Camino Bajo del Castillo s/n, Urb. Villafranca del Castillo,\\
~28691 Villanueva de la Ca\~{n}ada, Madrid, Spain\\
$^{15}$Institute of Cosmos Sciences, University of Barcelona, Mart\'i i Franqu\`es 1, 08028 Barcelona, Spain\\
$^{16}$Department of Physics, Durham University, South Road, Durham, DH1 3LE, UK\\
}

\date{Accepted XXX. Received YYY; in original form ZZZ}

\pubyear{2015}

\begin{document}
\label{firstpage}
\pagerange{\pageref{firstpage}--\pageref{lastpage}}
\maketitle

\begin{abstract}
We report on  the first simultaneous high-time resolution X-ray and infrared (IR) observations of a neutron star low mass X-ray binary in its hard state. We performed $\approx 2\,$h of simultaneous observations of 4U 1728-34 using HAWK-I@VLT, \xmm\ and \nustar.  The source displayed significant  X-ray and IR variability down to sub-second timescales. By measuring the cross-correlation function between the infrared and X-ray lightcurves, we discovered a significant correlation with an infrared lead of $\approx 30-40\,$ms with respect to the X-rays. {We analysed the X-ray energy dependence of the lag, finding a marginal increase towards higher energies. Given the sign of the lag, we interpret this as possible evidence of Comptonization from external seed photons.} We discuss the origin of the IR seed photons in terms of cyclo-synchrotron radiation from an extended hot flow. Finally, we also observed the IR counterpart of a type-I X-ray burst, with a delay of $\approx7.2\,$s. Although some additional effects may be at play, by assuming that this lag is due to light travel time between the central object and the companion star, we find that 4U 1728-34 must have an orbital period longer than $3\,$h and an inclination higher than 8$^\circ$.
\end{abstract}




\section{Introduction}

Low mass X-ray binaries (LMXBs) are systems in which a neutron star (NS) or a black hole (BH) is accreting mass from a small mass companion star. During their outbursts, they display a broad and complex spectral energy distribution which varies during different stages of their activity. Two main states can be identified \citep[see e.g.][]{Tananbaum1972,fender2004,migliari-fender2006,dgk07,lin2007atoll,belloni2011,munioz-dariaz2014}: a so-called \lq\lq soft state", in which the emission is mainly thermal, arising from an optically thick, geometrically thin accretion disk (dominating the soft X-ray part of the spectrum); and a ``hard state", in which  the X-ray spectrum is dominated by  a cut-offed power law, with a typical cut-off value of $\approx 10-20\,$keV for NSs and $\approx 100\,$keV for BHs  \citep[][]{burke2016}. This component is  explained in terms of  a geometrically thick, optically thin inflow (also referred to as "corona"), which comptonizes softer X-ray photons arising from the disk \citep[in BH;][]{dgk07,zdariskigierlinski} or the boundary layer  \citep[in NS;][]{lin2007atoll}. In the hard state, these systems can also show a flat spectrum which extends from radio to optical-infrared wavelengths, usually associated with synchrotron radiation from a compact jet \citep[see e.g. ][]{corbel2002,russell2007,migliari2010,baglio2016,diaz-trigo2018,tetarenko2021_a}. 

 It is generally accepted that the geometry of the accretion flow around accreting compact objects evolves during their outbursts \citep[see e.g.][]{esin1997,lin2007atoll,lin2009,ingram2010,demarco2017,wang-yanan-2017,wang-yanan-2019,vandeijnden_2020_1605,mendez2022}. However, the exact structure of the inflow and its connection to the jet is still a matter of debate \citep{kalemci2022}. A fundamental property of LMXBs that can help answering these key questions is the presence, during the hard state, of strong stochastic variability across the electromagnetic spectrum down to sub-second timescales \citep[see e.g.][]{motch1982,vanderklis1994,belloni2005,tetarenko2021_a}{}{}.   

 {The ultraviolet-optical-infrared (UV/O/IR) range contains contributions from three emitting components which vary along the outburst and act on different timescales: irradiated disk, jet and contribution from an extended hot inflow \citep{corbel2002,corbel2013,hynes2005,gierlinski,migliari2010,chaty2011,buxton2012, poutanen2014,kosenkov2020}. Thus, variability studies in these bands have proved to be extremely powerful to constrain the properties of LMXBs.}
 
 In recent years, the development of new fast O-IR photometers has allowed for a rapid growth of strictly simultaneous high-time resolution multiwavelength observations of BH LMXBs. The discovery of $\approx+0.1\,$s lag in at least 3 BHs (GX~339-4, V404~Cyg and MAXI~J1820+070) between X-ray and the O-IR emission clearly showed that mass accretion rate fluctuations can travel from the inflow to the jet, where they are re-emitted as synchrotron radiation \citep{casella2010,gandhi2010,gandhi2017,malzac2014,paice2019}.  This opened a new window in jet physics studies, providing new fundamental constraints on the jet speed \citep{casella2010,tetarenko_b_2018,Zdziarski-tetarenko2022}, the height and extension of the first shock region \citep[][]{gandhi2017,vincentelli2018,paice2019} and the launching radius \citep{vincentelli2019}.
 
 Further observations lead to the discovery of a non-linear correlation between the two bands \citep[][]{kanbach2001,durant2011,veledina2017,vincentelli2021_1535,paice2022}, as well as O-IR quasi-periodic oscillations \citep[][]{motch1983,kalamkar2016,vincentelli2019,vincentelli2021_1535}. This indicated the contribution of more than one variable component to the O-IR emission \citep[as also suggested by long term O-IR studies;][]{kosenkov2020}. One of the most promising candidates to explain these intriguing features is synchrotron radiation from the external regions of a hot magnetized inflow \citep[][]{veledina2011}.  Few LMXBs, including both BH and NS, have shown possible evidence of this component, but it is not clear under which conditions it can dominate over the jet \citep{degeneaar2014,veledina2017,paice2022,shahbaz2023}.

Despite the richness of O-IR phenomenology in BH LMXBs,  relatively few  studies focused on the fast multiwavelength properties of  NS LMXBs, mainly because of their lower luminosity\footnote{{The average X-ray peak luminosity of BH LMXBs is $\approx$7$\times$10$^{37}$ erg s$^{-1}$ against the $\approx$3.8$\times$10$^{37}$ erg s$^{-1}$ NS LMXBs \citep[][]{yan-yu2015}{}{}.  Furthermore,  stellar mass black holes are also brighter at O-IR wavelengths \citep[$L_{O-IR,BH}\approx\times$10$^{34}$erg s$^{-1}$, against $L_{O-IR,NS}\approx\times$10$^{33}$erg s$^{-1}$;][]{russell2006}{}{}.}}.   Due to their strong flaring and non-stationary variability, most of the attention focused on objects persistently accreting at high rates \citep[i.e. the so-called \lq\lq Z sources''; ][]{mcgowan2003,dubus2004,durant2011,shahbaz2023,vincentelli2023}, or transitional millisecond pulsars \citep{shahbaz2015,Shahbaz2018,papitto2019,baglio2019,baglio2023}. Yet, no study has been done for NS LMXBs in their  \lq\lq canonical'' hard state.

Here, we present the first simultaneous fast IR/X-ray observation of the accreting NS 4U 1728-34 (or GX 354-0, RA:  17 31 57.73, DEC:  -33 50 02.5) during its hard state. This system is a well known weakly magnetized neutron star (NS) {inferred to accrete from a H-poor donor} \citep{Shaposhnikov2003,galloway2010}. It is classified as a persistent atoll-type LMXB with high Galactic hydrogen column density, $N_{\rm H}= 2.4-4.5 \times 10^{22}$ cm$^{-2}$  \citep[][]{piraiano2000,dai2006,diaz-trigo2018}{}{}. 
Atoll-type systems are X-ray sources which can undergo  transitions between a soft and a hard state. When they do, they trace typical tracks in the X-ray colour-colour diagram (i.e. the so called \lq\lq Island\rq\rq\ and \lq\lq Banana\rq\rq\ branches), associated also to an evolution of timing properties \citep[see, e.g., ][]{hasinger1989,vanderklis2006}. Namely, as observed in BH LMXBs, the broadband noise components of the X-ray power spectrum, along with its quasi-periodic oscillations evolve towards high frequencies, as the source approaches a softer, higher luminosity phase \citep[see, e.g.][]{psaltis1999,homan2007,altamirano2008}. 
In detail, while the duration of these state can vary significantly from source to source \citep[see e.g.][ and references therein]{vanderklis2006,munioz-dariaz2014}, 4U 1728-34 undergoes regular state transitions every $\sim40-60$ days  {\citep[][ for previous detection of these periodicities see also: \citealt{kong1998,galloway2003}]{munioz-dariaz2014}{}{}}

Recent X-ray/IR observations in the soft state have detected the first infrared counterpart of a type-I burst \citep{vincentelli2020_burst}. These  rapid X-ray flashes of light arise from a thermonuclear runway on the surface of the NS \citep[][]{hansen1975,woosley-tam1976}. The low energy counterpart (detected at UV/O/IR wavelengths) of these events is usually associated to the reprocessed emission from the disk and the companion star. Thus,  the delay between the X-ray and the UV/O/IR emission can be used to constrain the orbital parameters of these systems \citep[see e.g.][]{hynes2006,munioz-dariaz2007,vincentelli2020_burst}.

\section{Data}

\subsection{Infrared: HAWK-I@VLT}

 We performed high time resolution photometric observations of  4U~1728-34 with HAWK-I \citep[][]{pirard}{}{}, mounted at the UT4 at the Very Large Telescope in Cerro Paranal, Chile. Observations were carried out  on the  2019-03-23 (MJD 58565) between 06:34 and 08:34 UTC in FastPhot mode (program ID: 0102.D-0182) using a window of $128\times64$ bins, allowing us to reach a time resolution of 0.125s. Data were recorded in N = 200 datacubes of 250 frames each, separated by $\approx3\,$s gaps. We then measured the background subtracted lightcurve from the target, a bright reference and a comparison star using the ULTRACAM software \citep[][]{dhilon2007}. In order not to lose the tracking of the stars, the relative position of the target and comparison star   was fixed with respect to the reference star. Within one cube the position of the aperture was able to tweak its position, following the stars. One set of aperture radii of the extraction was defined for each cube in order to take into account long term seeing variations. To remove any additional spurious effects, we then normalized the target lightcurve to the reference star ({VVV J173159.32-335001.92, K$_s=10.562\pm0.001$}). The lightcurve extracted from a nearby comparison star ({VVV J173200.33-335005.87, K$_s=14.58\pm0.04$}) showed a stable lightcurve without long term trends. The average magnitude of 4U 1728-34 was $14.5\pm0.02$ \citep[$\approx1.1\,$mJy, a similar flux  to past measurements by][]{diaz-trigo2018}. A fraction of the cubes suffered from a few frame losses. Therefore, when considering the cross-correlation with the X-rays, these cubes were discarded from the analysis. During a few cubes, the source ended up too close to the edge of the detector, distorting the obtained lightcurve: also these cubes were excluded from the overall analysis.  {After extracting the lightcurve, we put the times stamps in the Dynamical Barycentric Time (TDB) system using the JPL DE405 Earth ephemeridis adopting the method described in \citet{eastman2010}.}
\label{sec:maths}

 \subsection{Soft X-rays: XMM-Newton}
 We observed 4U 1728-34 with the EPIC-pn camera on board of the X-ray (0.5-10\,keV) telescope \xmm\ in Timing mode \citep{struder2001}. Observations were carried out on the same night between 05:37 and 13:32 UTC (\textsc{obsid: 0831791401}). After applying the barycentric correction (through the \textsc{barycen} tool {adopting the JPL DE405 Earth ephemeris}), we extracted the events between column RAWX 27 and 47, selecting only events with PATTERN<=4 and FLAG==0. The data also show a clear X-ray burst during the HAWK-I time window: for the analysis of the stochastic variability (and the correlation with the IR fast variability, see section \ref{sec:sto} and \ref{sec:en}), this feature was discarded from the data. In detail, we excluded events in a window between 4200 and 4400 s from the beginning of the observation.  The selected X-ray events were binned to the IR lightcurve.  Due to the very high count rate of this event, which distorts the shape of the lightcurve, for the analysis of the burst signal (see section \ref{sec:bur}) we obtained the lightcurve using the command \textsc{epiclccorr} with $0.5\,$s time resolution.

 \subsection{Hard X-rays: NuSTAR}

\nustar\ \citep{harrison2013} observed 4U\,1728--34 from the 2019-03-22 22:26:09 to 2019-03-23 12:11:09 UTC for a total on-source exposure time of 19.3\,ks (\textsc{obsid: 90501312002}) with both focal plane module A and B (FPMA and FPMB hereafter). We processed the event lists, filtered photons between 3 and 79 keV, excluded the passages of the satellite through the South Atlantic Anomaly using the tool {\tt nupipeline}. {Barycentric correction was also applied through the \textsc{ftool} command \textsc{baryocrr} (also in this case we used JPL DE405 Earth ephemeris)}. Both source and background counts were accumulated within a circular region of radius 150 arcsec. We then extracted the barycenter source light curves for both FPMs and inspected them for the presence of bursts. We detected one burst with a duration  $\approx100\,$s before the strictly simultaneous \xmm/HAWK-I campaign. The burst detected in the \xmm/HAWK-I campaign, instead, took place during a visibility gap. After excluding the identified burst, spectra and response files were generated applying the script {\tt nuproducts}. We binned the spectra to obtain at least a signal-to-noise ratio of 20.

 
\section{Data Analysis and Results}

\subsection{Stochastic variability}
\label{sec:sto}
\begin{figure}
\includegraphics[width=\columnwidth]{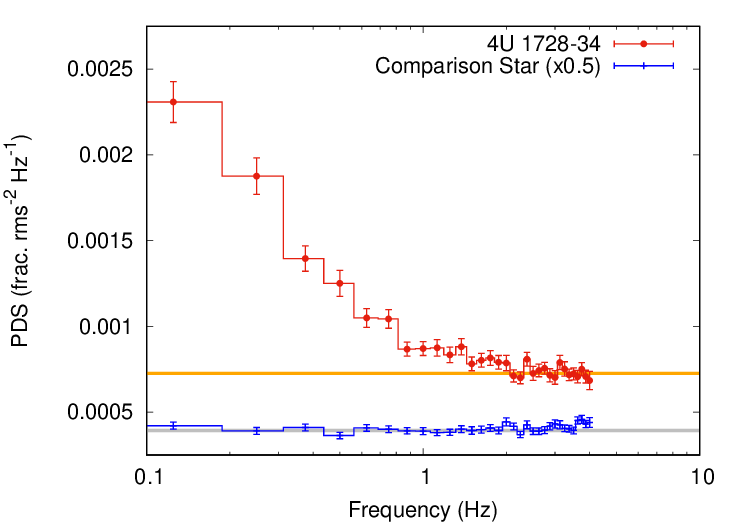}
\caption{IR Power Density Spectrum (PDS) of our target (4U 1728-34, red points)  and the chosen comparison star (blue points). While the former shows clear red noise, the latter is consistent with Poissonian noise.  The comparison star PDS has been rescaled for clarity. The grey line represents a constant fit to the comparison star PDS, {while the orange line represents the best constant fit in the range between 3 and 4 Hz.}} 
\label{fig:pds_hawki}
\end{figure}

Thanks to HAWK-I's subsecond time resolution we measured the Power  Density Spectrum (PDS) in the IR band using the recipe described by \citet{uttley2014}. We first investigated if the target was significantly variable computing the PDS with 64 bins {(i.e. segment length = 8s)} per segment for both our target and the comparison star (see Fig. \ref{fig:pds_hawki}). While the first shows clear red noise, typically observed in accreting systems \citep[see e.g.][and references therein]{press1978,uttley2005,scaringi2015}, the latter shows a flat spectrum, consistent with just uncorrelated noise. {Some deviations from the constant are seen (see e.g.  around 2 and 4 Hz). We note, however, that they lie within $2\,\sigma$ from our constant best fit, and thus they are unlikely due to instrumental artifacts that could affect the signal.}

\begin{figure}
\includegraphics[width=\columnwidth]{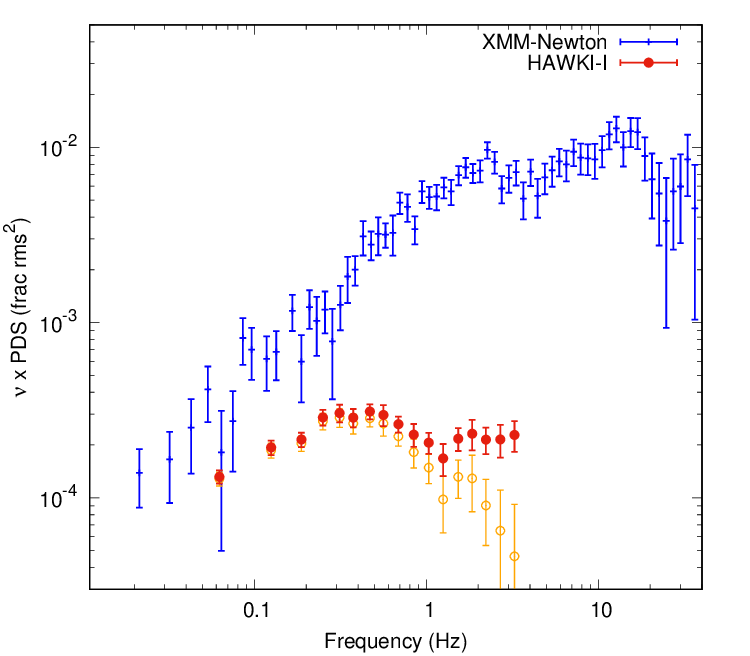}
\caption{ X-ray (blue points) and IR (red filled points) noise-subtracted PDS (in $\nu P_\nu$ units) of 4U 1728-34 during our observations. The yellow empty circles represent the IR PDS estimating the Poissonian noise fitting a constant between 3 and 4$\,$Hz{ (see orange line in Fig. \ref{fig:pds_hawki})}. The IR PDS has significantly lower variability and shows a break at lower frequencies.} 
\label{fig:pds_multi}
\end{figure}

We then computed a PDS with a better frequency resolution { and coverage (i.e. going at lower frequencies)}:  to do this we filled the gaps by adding log-normally distributed points which kept the same flux, and  high frequency PDS of our IR lightcurve. The PDS was computed using 128 bins per segment {(i.e. segment length = 16s)} and a geometrical rebinning of 1.2 (see Fig. \ref{fig:pds_multi}).   While at low frequencies the PDS shows a break around 0.3 Hz, it shows an upturn  above $\approx$ 1 Hz. Given the trend at frequencies below 1 Hz, such a change in slope is unlikely to be only aliasing. However, such an effect could be caused by an underestimation of the white noise level, which was computed following the recipe in \citet{vaughan2003}. Thus, we also computed the PDS by estimating the noise by fitting the constant level between 3 and 4$\,$Hz. Although this represents the most conservative estimate,  we still measure significant variability above 1 Hz (yellow empty points in Fig. \ref{fig:pds_multi}). The integrated rms computed in the former case is $\approx2\,$\%.  
  
We compared the IR PDS to the PDS of the \xmm\ data with 16348 bins per segment and a geometrical rebinning factor of 1.1. The X-ray PDS shows strong broadband noise  as typically observed in NS LMXBs in their hard state, with a peak (in $\nu P_\nu$ units) at around few Hz and an integrated rms of 20$\,$\%.  This value is consistent with typical expectations from the hard state \citep[][]{munioz-dariaz2014}{}{}.

\begin{figure}\includegraphics[width=\columnwidth]{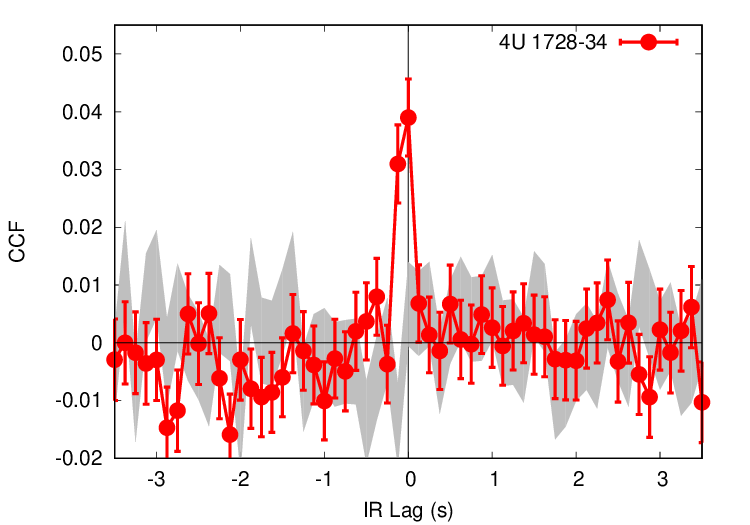}
\caption{Cross-correlation function (CCF) computed between \xmm\ and HAWK-I for 4U 1728-34 (red points) and the comparison star (grey curve).} 
\label{fig:ccf}
\end{figure}

We characterized the origin of this noise by measuring the discrete cross-correlation function (CCF) between the X-ray (0.5-10 keV) and IR using the same procedure described by \citet{gandhi2010}. As in all previous studies, we use the X-ray as a reference band: i.e. positive lag implies that the IR lags the X-rays. All plots follow this convention, unless specified otherwise. The \xmm/HAWK-I CCF has a poorly resolved peak around $\approx 0$, which seems to be slightly asymmetric around negative lags (Fig. \ref{fig:ccf}). By computing the lag as the weighted average of the cross-correlation lags between $\pm 0.5\,$s and propagating the errors accordingly, we obtain a lag of $-41\pm11\,$ms: i.e. IR \textit{leads} the X-ray variability. We compared this value with the Fourier lags: i.e. using the phase of the IR/X-ray cross-spectrum. This was computed following the steps described in \citet{uttley2014}.  {A comparable value is also obtained when computing the Fourier time lag between the X-ray and IR lightcurves at high frequencies: $-32\pm15\,$ms in the 1-4 Hz range and $-42\pm20\,$ms between 2 and 4 Hz.}

In order to verify the significance of the observed correlation, we performed two tests. We first evaluated the CCF between \xmm\ and the IR lightcurves of the comparison star: the absence of correlation shown in Fig. \ref{fig:ccf} (grey curve) demonstrates that the peak is not spurious. Second, we checked whether the peak was due to uncorrelated noise fluctuations. To do so, we simulated N = 10$^3$ synthetic lightcurves with the same PDS of the \xmm\ lightcurve and random phases, and cross-correlated them with the IR lightcurve. From the width of the obtained distribution we found that our peak is significant at $\approx 8\sigma$. {Finally, we also verified if the shape was affected by counting statistics, correcting the normalization from the excess variance \citep{vaughan2003}. However, we did not find a significant variation in the shape of the CCF, nor of its inferred lag.} Therefore we can conclude that the asymmetry of the CCF, and thus the IR lead, is real.

\subsection{CCF energy dependence}
\label{sec:en}

\begin{figure*}\begin{center}

\hspace{-0.1cm}\includegraphics[width=1.5\columnwidth]{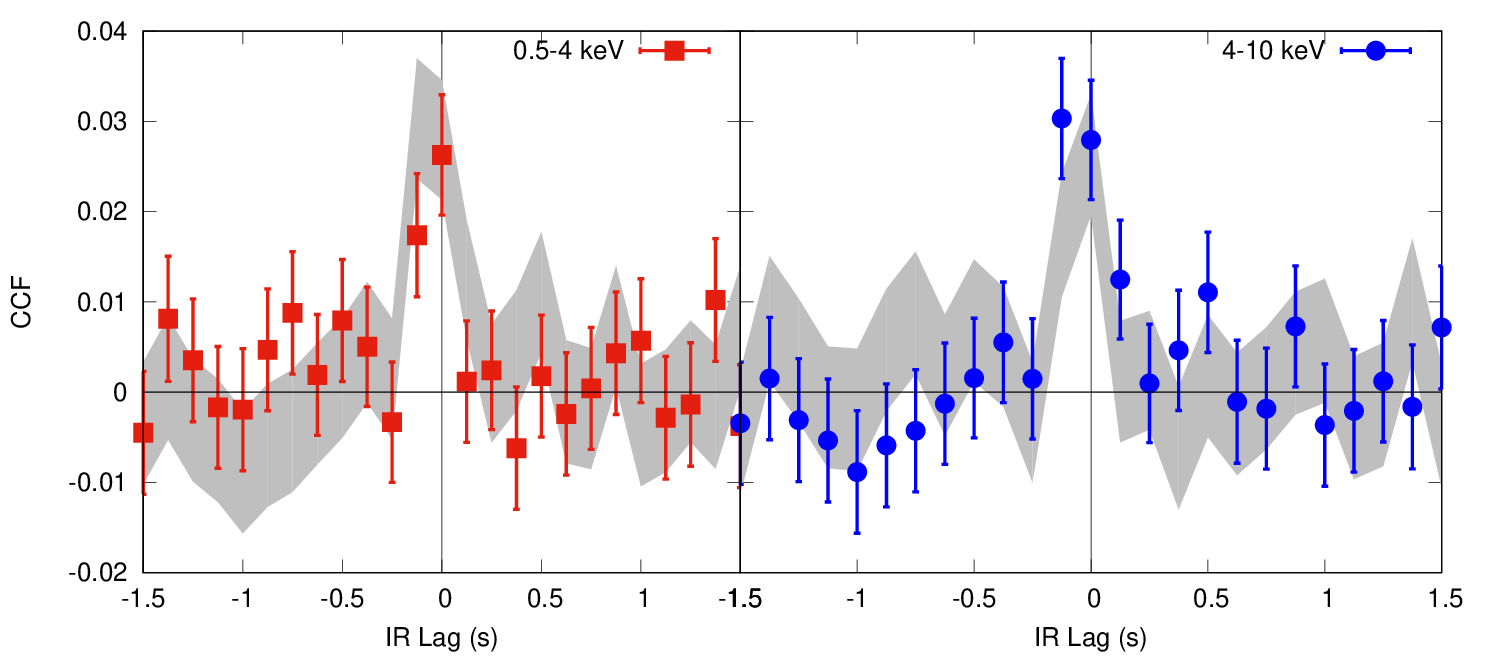}

\includegraphics[width=1.5\columnwidth]{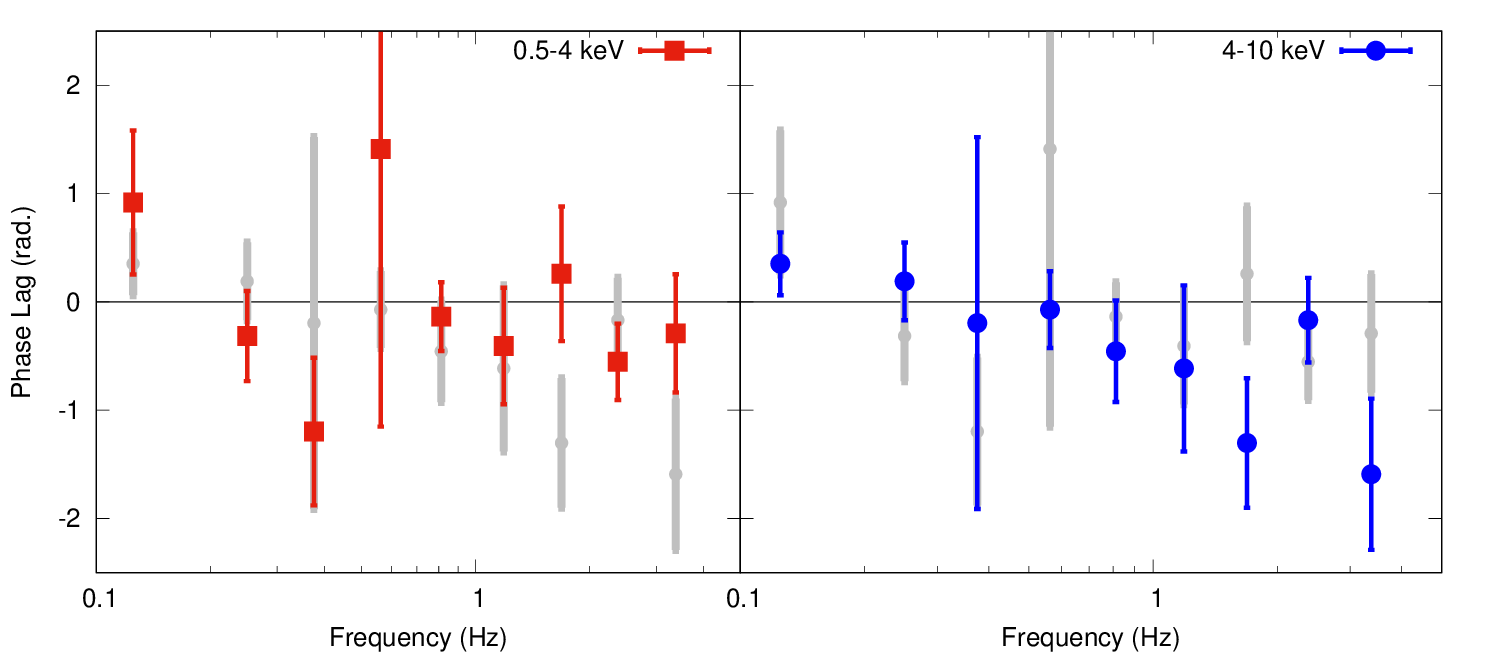}

\end{center}

\caption{ \textit{Top panel}: X-ray/IR CCF using different X-ray energies: 0.5-4$\,$keV (red squares on the left) and 4-10$\,$keV (blue circles on the right).  \textit{Bottom panel}:  X-ray/IR phase lags using the same bands (and color code). To ease comparison, on the left panels grey lines represent measurements done with the 4-10$\,$keV band and on the right panels grey lines reproduce the calculations using the 0.5-4$\,$keV band.} 
\label{fig:lag_vs_energy}
\end{figure*}

We investigated the CCF energy dependence by using \xmm\ lightcurves extracted in the  0.5-4 and 4-10$\,$keV bands. The resulting plots (see Fig. \ref{fig:lag_vs_energy}, Top panels) show a possible evolution between the two bands. To exclude that this is an effect due to the normalization, we evaluated the Fourier domain phase lags between the IR and the two X-ray bands  by using  64 bins per segment and a geometrical rebinning factor of 1.4. Although the statistics is poor, the bottom panels of Fig. \ref{fig:lag_vs_energy} show that the resulting lag-frequency spectra appear to  marginally deviate from each other above 1$\,$Hz (where the lag in the  4-10$\,$keV band becomes longer). 

\begin{figure}\begin{center}

\includegraphics[width=\columnwidth]{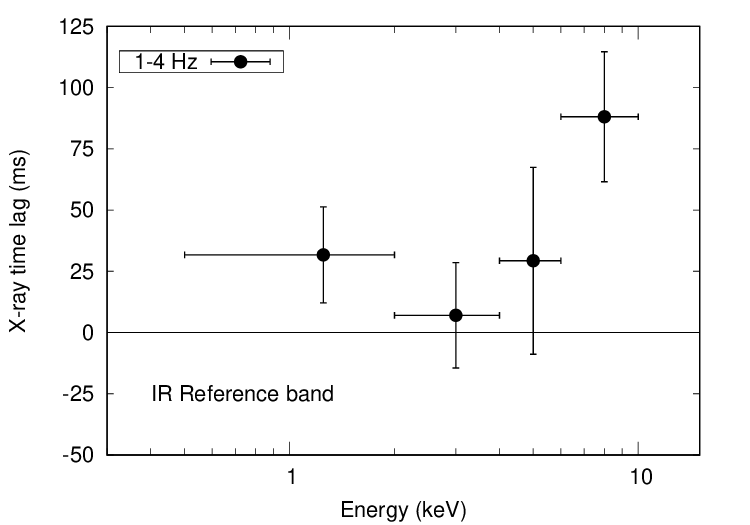}
\end{center}

\caption{Lag-energy spectrum using the IR as a reference band: i.e. positive lag means X-ray lagging the IR. The plot shows how the lag seems to become larger at energies above 6$\,$keV.   } 
\label{fig:lag-spec}
\end{figure}

In order to {investigate further} this result, we computed the lag-energy spectrum by integrating the cross-spectrum between 1 and 4$\,$Hz for four X-ray energy bands (0.5-2, 2-4, 4-6 and 6-10$\,$keV). Given the IR lead, we used the IR bands as reference. We caution the reader that this  \textit{inverts} the convention used so far: i.e.\textit{ positive lag means that the X-rays lag the IR}. The resulting time-lag vs energy-spectrum is shown in Fig. \ref{fig:lag-spec}.   While the IR lead (or X-ray delay) does not evolve significantly between 0.5 and 6$\,$keV, a marginal increase can be seen in the 6-10$\,$keV band. {We also attempted to measure the lags using broader bands (0.5-4 and 4-10 keV, i.e. as for the analysis shown in Fig. \ref{fig:lag_vs_energy}) and we found only a marginal a difference of $\approx$ 1.5 $\sigma$: 17$\pm$15 ms to 62$\pm$27 ms}. We note that, as  expected from the lag-frequency spectrum, by performing the same analysis at lower frequencies (i.e. 0.1-1$\,$Hz), all lags are consistent with zero.

  \subsection{Type-I burst}

\label{sec:bur}

\begin{figure*} 

\hspace{-0.1cm}\includegraphics[width=\columnwidth]{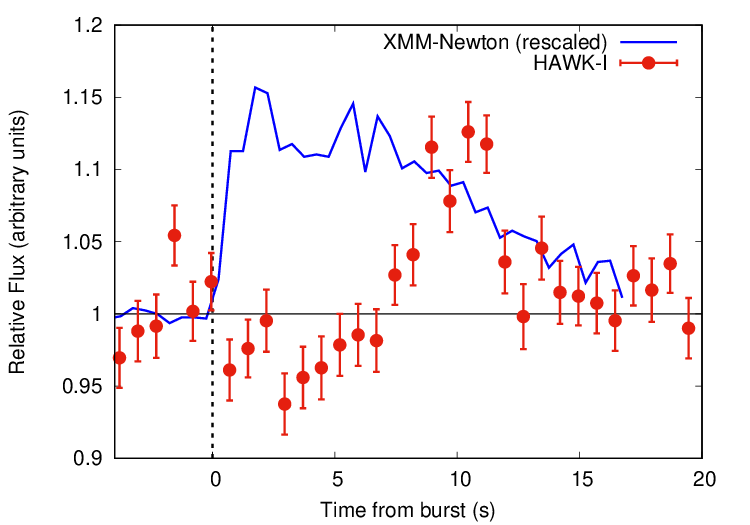}
\hspace{-0.1cm}\includegraphics[width=\columnwidth]{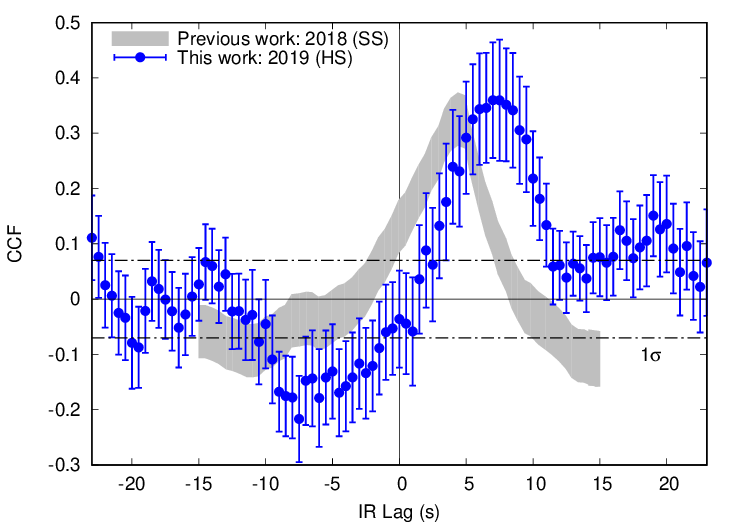}

\caption{ \textit{Left Panel:} \xmm\ (blue line) and HAWK-I lightcurve (red points, rebinned to $0.75\,$s for clarity) during the type-I burst. The dashed line was placed at the rise of the burst (i.e. between the first bin with a clear rise, and the last consistent with the persistent). The IR lightcurve shows a marginal anti-correlation around the peak of the X-ray burst followed by a clear rise. The gap at the end of the \xmm\ burst is due to telemetry drop-out. \textit{Right~Panel:} CCF of the hard state  (HS, blue points) compared with the one measured in the soft state \citep[SS, grey curve; ][]{vincentelli2020_burst}. The latter was computed only up to a lag of $\pm15\,$s because of the limited duration of the simultaneous data. Nonetheless, a clear difference between the two epochs is present. {The  dashed-dotted lines represent the 1$\sigma$ confidence noise level for the hard state observations.}} 
\label{fig:ccf_burst}
\end{figure*}

As mentioned above, the \xmm\  lightcurve presented a clear X-ray burst simultaneous with the HAWK-I cubes. However, due to the presence of stochastic variability, the IR lightcurve does not show a clear excess as in the previous burst reported by \citet{vincentelli2020_burst}.  Nonetheless, by measuring the CCF including the burst, we found a broad peak which suggests that the IR does respond to the X-ray burst also in the hard state (see Fig. \ref{fig:ccf_burst}, right panel). Interestingly, the CCF seems also to show a  weak anti-correlation at $\approx-5\,$s. By inspecting the IR lighcurve with lower time resolution, we found the presence of  a marginal dip  during the initial phase of the burst followed by a positive response to the X-ray burst (see Fig. \ref{fig:ccf_burst}, left panel), which could explain the weak anti-correlation at negative lags. {We estimated the significance of this correlation by cross-correlating the XMM lightcurve including the burst with an uncorrelated signal with the same properties of the IR lightcurve. We simulated $10^3$ lightcurves with the same signal to noise ratio and power spectrum measured by HAWK-I, aligned them to the XMM lightcurve and computed the CCF for all the possible shifts. The $1\,\sigma$ threshold obtained from the distribution is plotted in Fig.\ref{fig:ccf_burst}, (right panel).  The peak of the CCF is highly significant ($\approx 4.4\,\sigma$), while, as expected, the dip at negative lag is just around $\approx 3\,\sigma$.}

Following the procedure described by \citet{gandhi2017} and \citet{peterson1989}, we evaluated the X-ray/IR delay through the weighted average (and its standard error) of the CCF  within its full width half maximum, finding a lag of $7.2\pm1.4\,$s. Past observations in the soft state showed a delay associated to the burst of $4.5\pm0.25\,$s \citep[][]{vincentelli2020_burst}, thus smaller with a $\approx 2 \sigma$ confidence level.

\subsection{X-ray spectral analysis}

\begin{figure}\begin{center}

\includegraphics[width=\columnwidth]{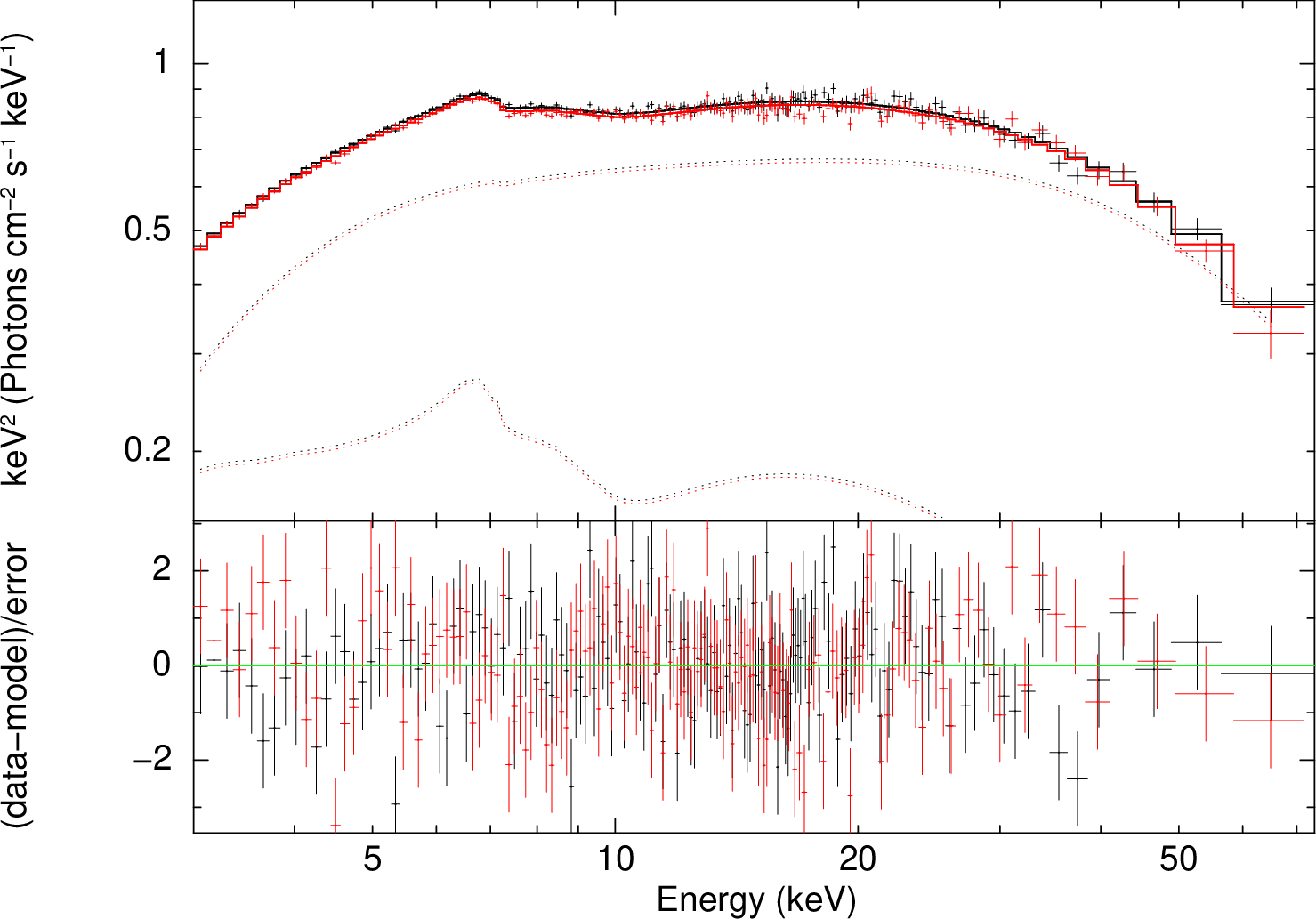}
\end{center}

\caption{ \textit{Top Panel:} \nustar\ spectra of 4U 1728-34. Black and red points represent the data from FPMA and FPMB detectors respectively. The continuous line represents the overall model, while the dashed lines indicate the individual components. \textit{Bottom Panel:}  Residuals of the fit. } 
\label{fig:spec}
\end{figure}

We also characterized the X-ray spectral state of the source, by analysing the \nustar\ spectrum between 3 and 79$\,$keV (Fig. \ref{fig:spec}). Following past spectral studies of NS LMXBs \citep[][]{wang-yanan-2017,wang-yanan-2019,ludlam2019,vandeijnden_2020_1605}{}{}  we used a model with four components: interstellar absorption, a black body component, a cut-off power law and reflection. In details, we used a  Comptonization model \citep[\textsc{thcomp};][]{Zdziarski2020} convoluted with a black body component   (\textsc{bbodyrad}) and a relativistic reflection model \citep[\textsc{relxillCp};][]{garcia2014}. We fixed the hydrogen column density N$_{\rm H}$ to the value reported in the \textsc{heasarc}\footnote{\url{https://heasarc.gsfc.nasa.gov/cgi-bin/Tools/w3nh/w3nh.pl}} database for the source coordinates, i.e. $1.35\times10^{22}\,$cm$^{-2}$ \footnote{{Past studies have found a better spectral fit by leaving the asborption free, finding $N_H$ $\approx 2.5\times10^{22}\,$cm$^{-2} $\citep{piraiano2000,dai2006,diaz-trigo2018}. We note that we also performed the analysis with this value and (probably due to the energy band pass) did not find any significant variations in the inferred parameters.}}.

The model can reproduce with good agreement the data ($\chi^2/dof\approx1.07$) and the best fit values are reported in  Table \ref{tab:fit}. We found the presence of a thermal component at  $kT \approx 1\,$keV, thus indicating  most probably the boundary layer \citep[see e.g.][]{lin2007atoll,armaspadilla2017}{}{}. The temperature of the hot Comptonizing medium was found to be $kT_e \approx15\,$keV. Both these values are consistent with past measurements of NS LMXBs X-ray spectra during the hard state \citep[][]{lin2007atoll,burke2016}{}{}. The value of the covering fraction is consistent with 1, which indicates that all the seed photons are Comptonized. Regarding the reflection component, the spectrum is consistent with a highly ionized, ($\log \xi \approx 3.5$), truncated ($R_{in}\approx10^7$cm) disk with a low inclination ($i$ between $\approx20^\circ$ and $30^\circ$). The reflection model allows us to define two disk emissivity profile indices and a radius for the transition between the two indices. Since the fit does not require strong relativistic effects (the disk is truncated), we fixed both indices to 3 which is what is predicted in the Newtonian case. The spin   was also fixed to 0, for the same reason. Furthermore, by leaving the density ($\rho$) of the disk as a free parameter, we find a value of $\rho\approx 2.5\times10^{18}\,$cm$^{-3}$.

{As mentioned above, this observation has a much harder spectrum compared to the first burst detection reported in \citet[][]{vincentelli2020_burst}{}{}. In particular, the previous observation is dominated by a strong thermal component, which can be fitted with a disk black body emission with a temperature of $\approx$3 keV. We also find evidence of reflection features as well as a hard tail above 50 keV. This is in line with the previous studies of the source \citep[see e.g.][]{kakava2017,wang-yanan-2019}{}{}, and a more detailed analysis on the evolution of the spectrum, beyond the aim of this paper, is ongoing (Vincentelli et al., in prep.).  }

\begin{table}

\caption{Best fit parameters for the \nustar\ FPMA + FPMB spectral analysis. The data were well reproduced assuming an absorbed Comptonization model + reflection: specifically, we used the model \textsc{phabs} $\times$ (\textsc{thcomp} $\circledast$ \textsc{bbodyrad} + \textsc{relxillCP}). N$_H$ was fixed to $1.35\times10^{22}\,$cm$^{-2}$, $a=0$ and both emissivity profile indices to 3. All errors are reported with 90$\,$\% confidence level. }
\centering
\begin{tabular}{lllll}
\hline
Component          & Parameter         &  & Value               \\
\hline

\textsc{thcomp}             &                   &  &                     \\
                   & $\Gamma_\tau$         &  & 1.9$\pm$0.01           \\
                   & $kT_e$ (keV)              &  & 15.4$^{+0.3}_{-0.6}$      \\
                   & $cov_{frac}$          &  &0.99$^{+0.01}_{-0.09}$ \\

\textsc{bbodyrad}           &                   &  &                     \\
                   & $kT$ (keV)             &  & 0.96$^{+0.07}_{-0.01}$      \\
                   & norm              &  &   154$^{+28}_{40}$         \\

\textsc{relxillCp}           &                   &  &                     \\
                   & inc ($^\circ$)              &  & 23$^{+13}_{-9}$        \\
                   & R$_{in}$  ($\times$6 GM/c$^2$)             &  &  6$^{+9}_{-2}$              \\
                   & log$\xi$             &  & 3.5$\pm$0.1            \\
                   & log $\rho$   (cm$^{-3}$)          &  &  18.4$^{+0.7}_{-0.5}$       \\
                   & norm ($\times$ 10$^{-3}$) &  & 1.8$\pm$0.4            \\
                   \hline

$\chi^2$ /dof (FMPA)  &                   &  & 491 /     485       \\
$\chi^2$ /dof (FMPB)                    &                   &  & 546   / 466      \\
$\chi^2$ /dof (total) &                   &  & 1037 / 961 & \\
\hline

Flux 3-79 keV\space(erg cm$^{-2}$ s$^{-1}$) &                   &  & 3.6 $\times$10$^{-9}$ & \\
Flux 3-6 keV\space\space\space(erg cm$^{-2}$ s$^{-1}$) &                   &  & 7.3 $\times$10$^{-10}$ & \\
Flux 6-9 keV\space\space\space(erg cm$^{-2}$ s$^{-1}$) &                   &  & 5.5 $\times$10$^{-10}$ & \\
\hline

\end{tabular}
\label{tab:fit}
\end{table}

\section{Discussion}

\subsection{Stochastic variability and lags}
Our strictly simultaneous \xmm/HAWK-I observations of 4U 1728-34 show a strong correlated signal with an infrared lead of $\approx40\,$ms. Given that the CCF structure is unresolved it is not straightforward to give a physical explanation to this delay. {Moreover, it is the first time that the sub-second IR domain is explored for a weakly magnetized NS LMXB in the hard state. So, there are currently no models which can predict this feature. Given the similarity between BH LMXBs and these systems \citep[see e.g.][]{psaltis1999,munioz-dariaz2014,vincentelli2023}{}{}, we attempted to interpret this signal through models recently developed for stellar mass black holes.}

{Historically, broadband spectral studies of LMXBs have usually interpreted the presence of an IR excess in the hard state as a jet component \citep{corbel2002,russell2006,migliari2010,gandhi2011,corbel2013,baglio2016,diaz-trigo2018,marino2020}.  From a timing point of view, this scenario  has been confirmed only for BH LMXBs thanks to the detection of a 0.1s lag between the X-ray and O-IR variability \citep[][]{gandhi2008,casella2010,malzac2014}{}{}. According to this model, accretion-rate fluctuations from the inflow (which are observed in X-rays) are injected in the jet, where they are re-emitted in the form of synchrotron radiation at lower energies (from O-IR to radio) through the collisions of internal shocks \citep[][]{jamil2010,malzac2013}{}{}: therefore the IR responds to the variations in X-rays with a positive lag. }

{The IR lead found in our analysis seems to disfavour a pure jet model. However, it is important to recall that due to the poor resolution of the CCF, we cannot rule out completely the presence of a jet. A component with a (positive) lag, smaller than half the time resolution of our CCF, could still be present. On this regard, it is also interesting to notice that the X-ray/IR delay from internal shocks is expected to scale with the mass of the central object \citep{malzac2013}. Thus, if the jet properties are similar between BHs and NSs, we expect a lag associated to the jet of $\approx 10\,$ms.  Our observations set an upper limit to this lag of $\approx 60\,$ms (i.e. half the time resolution of the CCF). This scenario can be tested with higher time resolution IR instruments \citep[see e.g. ERIS@VLT][]{erisdavies}{}{}.}

{Another scenario which has been invoked to explain the O-IR flux excess in the hard state, is the production of cyclo-synchrotron radiation from an extended hot inflow \citep[][]{veledina2011,veledina2013}{}{}. According to these models, the X-rays would be then produced by the Comptonization of these photons \citep{wardzinskyzdziarski2000,poutanen2009}. The sign of the lag, seems to suggest that Comptonization may be involved. Depending on the energy distribution of the electrons and/or the variability of some parameters of the system, a positive correlation with an O-IR lead can be reproduced \citep{poutanen2009,veledina2017,paice2022}. }  

{In the following paragraphs we discuss the possible implications regarding a cyclo-synchrotron emitting region. These calculations need the value of the IR luminosity, which depends on distance and extinction. Regarding the distance, we  will use the value  obtained through photospheric radius expansion burst measurements \citep[i.e. $\approx 5\,$kpc;][]{galloway2003}. The extinction is more uncertain as there is no direct measurement of the $N_{\rm H}$. For the spectral analysis, we found a good fit using $1.35 \times 10^{22}\,$ cm$^{2}$ (see Table \ref{tab:fit}), however past works have also found good fits using a higher $N_{\rm H}$ (roughly by a factor of 2). Using the relation found by \citet{guver2009} such values of $N_{\rm H}$ lead to  A$_V$ ranging between $\approx 6$ and 11. Following \citet{cardelli1989} law, these in turn  translate into A$_K$ between $\approx$ 0.7 and 1.3. For our calculations we kept consistent with our spectral fit, and adopted the lower value of A$_K$. We caution, however, that the discussed values may change by a factor of a few depending on the extinction.
}

If the IR emission coincides with the peak of the cyclotron spectrum, it is possible to constrain the geometry of the emitting region. In details, adapting the equations described in \citet{masters1977} for accreting white dwarfs,  the radius of a slab emitting cyclotron radiation can be written as:

\begin{equation}
R = \sqrt{ \frac{12c^2}{32\pi^2} \frac{L_{cyc }}{kT \nu^3}} 
\label{eq:cycl}
 \end{equation}

where $L_{cyc}$ is the observed luminosity at the electromagnetic frequency $\nu$ ($1.5\times10^{14}\,$Hz), and $T$ is the temperature of the slab. { The exact temperature of the cyclotron emitting region is not known. However, we can assume that its temperature will not be higher than the coronal temperature inferred from the X-ray spectral fit ($kT_e\approx$15keV). Therefore this will give us a lower limit to the radius. We note that this is valid under the  assumption of a thermal Maxwellian distribution for the electrons, and the value of the effective temperature could change depending on the energy distribution of the electrons.} We obtain a de-reddened K$_S$ band luminosity $L_{cyc}$ of $\approx10^{33}\,$erg s$^{-1}$; given, however given an IR rms of $\approx$ few percent, the luminosity of the variable component (responsible for the CCF) could be lower as $\approx10^{31}\,$erg s$^{-1}$. By plugging in these values in eq. \ref{eq:cycl}, we find that $R$ has to be {at least of the order of} $10^8-10^9\,$cm. Such values seem unlikely, as they would be marginally consistent with the light travel time distance between the X-ray and the IR emitting region ($\approx10^{-2}$ light seconds, i.e. $10^8$\,cm).

{We also considered the case of an IR emitting synchrotron region. In detail, similarly to  \citet{chaty2011},  we used a one-zone cylindrical region emitting IR synchrotron radiation. Although a more realistic description should include a stratified hot flow \citep[][]{veledina2013}{}{}, we adopt this approximation as it allows us  to find a relatively simple relation between the fundamental quantities such as radius ($R$) and height of the slab ($h$), magnetic field ($B$), break of the synchrotron radiation and flux. Thus, we can constrain the parameters of the system as well as the geometry of the corona.}  In details, assuming equipartition ($\xi$=1), and the distance of 4U~1728-34, equations 1 and 2 from  \citet{chaty2011} can be written as follows:

\begin{equation}
B = C_B\nu\left(\frac{h}{S}\right)^\beta [\rm{G}]
\label{eq:bsync}
 \end{equation}

\begin{equation}
R = \frac{C_R}{\nu}\left(\frac{S}{h}\right)^\alpha [\rm{cm}]
\label{eq:rsync}
 \end{equation}
where $\nu$ is the break of the synchrotron spectrum expressed in units of $10^{14}\,$Hz, $S$ is the flux in units of 10 mJy, $h$ is the height expressed in units of $R$ (i.e., $H/R$), while $C_R$ and  $C_B$ are constants (respectively $2.5\times10^8$ and $5\times10^5$). The values of the indices $\alpha$ and $\beta$ are respectively 17/36 and 1/9. 

 \citet{diaz-trigo2018} measured the broadband spectral energy distribution of 4U~1728-34 in a similar spectral state, finding a break frequency between $1.3 - 11.0 \times10^{13}\,$Hz ($\nu \approx 1.3 - 11.0 \times 10^{-1}$).  Moreover, given the sign of the lag it is implausible that most of the variability arises from the jet. Thus, we can assume that the  height of the cylinder is smaller than the radius (i.e. $h_{max}=1$). Finally, given an rms of a few percent, we can place a lower limit to the (variable) synchrotron flux of $\approx10^{-2}\,$mJy ($S_{min} \approx 10^{-3}$). By combining these three quantities we can constrain the magnetic field and radius of the slab. For the former, this translates into:

\begin{equation}
B < B_{max} = C_B\nu_{max}\left(\frac{h_{max}}{S_{min}}\right)^\beta \approx 10^6 \,\rm{G}
\label{eq:bsyncmax}
 \end{equation}

Such a value is in line with the expectations from weakly magnetized sources, such as 4U~1728-34 \citep[see e.g.][]{casella2008,ludlam2019}{}{}. When considering the radius, instead, we obtain:

\begin{equation}
R > R_{min} = \frac{C_R}{\nu_{max}}\left(\frac{S_{min}}{h_{max}}\right)^\alpha \approx 9\times10^6 \,\rm{cm}
\label{eq:rsyncmin}
 \end{equation}

Such a value is consistent with the value inferred through our X-ray spectral fitting ($R_{in}=5-19 \times10^6\,$cm). Thus, if we now consider the upper limit placed by the \nustar{} data analysis (i.e. fixing $R=R_{max}=1.9\times10^7$ cm) we can also rearrange the equations and place constraints on $h_{min}$ and $\nu_{min}$. Regarding $h$ we obtain:

\begin{equation}
h > h_{min} = \left(  \frac{C_R}{\nu_{max}R_{max}}\right)^\frac{1}{\alpha} S_{min} \approx 0.2
\label{eq:hsync}
 \end{equation}
which would point towards a thick wedge.  We note that recent X-ray polarimetric measurements on other low magnetic field NS LMXBs at higher accretion rates have pointed towards a spherical geometry, with a smaller truncation radius \citep{capitanio2022,chattarjee2023-gx9+9,farinelli2023}. Thus, the lack of fast IR stochastic variability in the soft states \citep{vincentelli2020_burst} could be explained with a contracting hot flow as the source approaches higher accretion rates \citep[see e.g.][]{esin1997,wijnands-vanderklis1999,ingram2010,munioz-dariaz2014,vandeijnden_2020_1605}. 
 More data of these sources with both fast multiwavelength timing and polarimetric observations may thus be used in the future to place stronger constraints on the evolution of this component as a function of luminosity. 

Regarding the synchrotron break we obtain:

\begin{equation}
\nu > \nu_{min} = \frac{C_R}{R_{max}}\left(\frac{S_{min}}{h_{max}}\right)^\alpha \approx  0.5 ~~~~~ [\times10^{14} \,\rm{GHz}]
\label{eq:nusync}
 \end{equation}

This is  consistent with the results obtained by \citet{diaz-trigo2018} and suggests that the turn-over between optically thin and optically thick region is in the near-IR band. New multiwavelength observations, with mid-IR coverage are required to confirm this scenario.

\subsection{Type-I burst}

We have measured for the second time the delayed IR counterpart of an X-ray type-I burst in 4U 1728-34. It is generally accepted that these X-ray flashes occur because of a thermonuclear runway on the surface of the accreting NS \citep[][]{hansen1975,woosley-tam1976}.  The presence of UV/O/IR counterparts in these systems is usually interpreted as reprocessing from the outer disk and the companion star \citep{obrien,hynes2006}.

Interestingly, we found a variation in the X-ray/IR delay, passing from 4.5 to 7.2$\,$s compared to a previous observation \citep{vincentelli2020_burst}. Such a marked difference suggests that the IR burst originates from thermal reprocessed emission of the donor star. If this is due only to a change in the orbital phase (i.e. the change of hot inflow geometry does not affect the delay between the driving and reprocessed burst), this measurement can be used to further constrain the period of the system and put a limit on the inclination.

\subsubsection{Geometrical interpretation: limits on the orbital period}

 Similarly to \citet[][]{vincentelli2020_burst}, by simply considering the donor star as a point source and assuming that the 7.2s is measured at superior conjunction with $i=90^\circ$, we can set a lower limit to the orbital period.  In that case the delay ( $\tau$) would correspond to $\tau=2a/c$  (where $a$ is the orbital separation and $c$ is the speed of light), plus the reprocessing time \citep[which is expected to be less than a few 100$\,$ms;][]{cominsky1987}{}{}. Even considering a more realistic case, where the star occupies a significant fraction of the orbital separation, or if the IR emission arises from a large fraction of the star, this limit would still hold. Thus, given a delay of 7.2$\,$s, we can say that $a=3.5c$.   According to Kepler's third law (for a canonical neutron star mass of 1.4$\,$M$_\odot$), this yields a minimum orbital period $\approx3\,$h, ruling out the ultra-compact nature of the source proposed by \citet{galloway2010}. This would also be in line with the period derived from recent transient pulsations searches \citep[$P_{orb}$ between $2\,$h and $8\,$hr;][]{bahar2021}.

\subsubsection{Geometrical interpretation: limits on the inclination}

Under the hypothesis that the IR burst arises from reprocessing, the measurement of the delays from two different events can be used to put the first lower limit to the inclination of the source. The light travel time distance between the central and donor star in an orbiting system with  a semi-major axis $a$, with an inclination $i$ at an orbital phase $\phi$ is \citep{obrien}:

\begin{equation}
    \tau= \frac{a}{c} \cdot [1 + \sin i \cdot \cos \phi]
        \label{eq:lag1}
\end{equation}

Thus, if we have two bursts of the same system  measured at different phases, we will observe a difference in the lag between the driving and reprocessed burst. We can express the difference between the measured reprocessed lags as:

\begin{equation}
\centering
    \Delta\tau = \tau_2 -\tau_1 = \frac{a}{c} \sin i \cdot [ \cos\phi_2-\cos\phi_1] 
    \label{eq:lag2}
\end{equation}

For a given inclination, the maximum observable time difference between two delays will be seen when the phases of the two bursts are 0 and $\pi$, thus:

\begin{equation}
\centering
    \Delta\tau < \frac{2a_{\rm{max}}}{c} \sin i
\end{equation}

By assuming a maximum $a_{\rm{max}}/c=10$ (i.e. the value of semi-major axis for an orbital period of $\approx20\,$h \footnote{The the maximum orbital period which can be sustained with a stable outburst according to the disk-instability model for 4U1728-34's mass accretion rate is $\approx10$-$20\,$h \citep{dubus_2019}.}), this gives $i>8^\circ$. This is consistent with previous modelling of the source \citep{vincentelli2020_burst} and sets the first geometrical constraint on the inclination of 4U 1728-34.

\subsubsection{Additional effects}

It is interesting to notice that the observed IR burst shows also some properties that are not fully consistent with the reprocessing scenario. Its duration is significantly shorter than that of the X-ray burst, and the shape does not fully resemble the observed one at high energy \citep[opposite to  the soft state case;][]{vincentelli2020_burst}. The shorter duration could be explained by assuming a high continuum level \citep[see e.g.][]{vincentelli2020_burst} and that we observe only a small fraction of the star reprocessing the X-rays \citep[i.e. if the star is between us and the NS;][]{obrien}. A different reprocessed burst profile, however, could also arise if the burst observed by the outer regions of the system is different from the observed X-ray burst. A possible explanation for this could come from the effect of an extended hot flow, which is present during this state \citep[a similar argument has been invoked also to explain the lack of correlation between AGN X-ray and optical/UV lightcurves. See e.g.][]{gardner2017}. For example, if the corona is depleted during the type-I burst (as suggested by high energy observations and simulations) and reappears during its decay, it may scatter the X-ray flux in a complex way, partially obscuring the NS surface to the companion. This would lead to a "truncated" driving signal, explaining the different shape of the reprocessed burst.

An additional signature of the corona is the presence of a marginal \lq\lq dip" in the IR lightcurve at the peak of the X-ray burst. Further observations are required to  confirm this feature, however, it is interesting to note that such an anti-correlation would be expected if the IR arose from synchrotron radiation by the corona \citep{degenaar2018}. If the corona cools down due to the dramatic increase of soft seed photons from the thermonuclear flash \citep[as supported by high energy observations of type-I X-ray bursters, see e.g.][]{maccarone-coppi2003,chen2013,hmxt2018_burts1636,Ji2013,ji2014,ji2014ab,kakava2017,sanchez2020}, then the synchrotron radiation is also expected to  then the synchrotron radiation is also expected to drop. Consolidation of this dip with further observations may lead to new independent constraints on the evolution of the accretion flow during these extreme events.

 \section{Conclusions}
 
We have performed the first simultaneous, sub-second IR/X-ray observations of an accreting neutron star in the low hard state. We found two main results:

   \begin{itemize}
   \item We  discovered a correlated signal between X-ray and infrared on millisecond timescales with an IR lead of $\approx30-40\,$ms.  There are no specific models that can explain this feature. However, given the energy dependence of the lag, we propose a scenario in which IR cyclotron or (more likely) synchrotron photons are produced by an extended hot inflow, and then Comptonized by the more energetic electrons responsible for the X-ray non-thermal emission. 
\smallskip

   \item We detected the IR counterpart of a type-I burst with lag of $\approx 7.2\,$s with respect to the X-ray one. By combining this result with past measurements, and assuming that the change is not affected by the spectral state,  we obtain a minimum orbital period of $\approx3\,$h and a lower limit to the inclination of $\approx8^\circ$. However, we also note that the shape of the IR burst suggests that some interaction with the corona is at play.
\end{itemize}
These results confirm the strength of fast multiwavelength variability for studying accretion processes. Moreover, the  discovery of correlated X-ray/IR subsecond variability 4U 1728-34 opens a new window in the study of NS LMXBs. Further observations with new IR instruments with higher time resolution \citep[e.g. ERIS@VLT;][]{erisdavies} may be able to resolve the observed peak in the CCF, leading to unprecedented constraints on the geometry of the accretion flow and jet's physical conditions in NS LMXBs. 

\section*{Acknowledgements}

{We thank the referee for the valuable comments which improved the quality and readability of the paper.
}This work benefited from the discussions done during the ISSI meeting in Bern "Looking at the disc-jet coupling from different angles".
FMV acknowledges support from the grant FJC2020-043334-I financed by MCIN/AEI/10.13039/501100011033 and Next Generation EU/PRTR. This work is supported by the Spanish Ministry of Science under grants PID2020–120323GB–I00, PID2021–124879NB-I00, and EUR2021–122010 
TMB acknowledges financial contribution from grant PRIN INAF 2019 n.15. YC acknowledges support from the grant RYC2021-032718-I, financed by MCIN/AEI/10.13039/501100011033 and the European Union NextGenerationEU/PRTR. LS acknowledges financial contributions from ASI-INAF agreements 2017-14-H.O and  I/037/12/0; from “iPeska” research grant (P.I. Andrea Possenti) funded under the INAF call PRIN-SKA/CTA (resolution 70/2016), from PRIN-INAF 2019 no. 15 and from the Italian Ministry of University and Research (MUR), PRIN 2020 (prot. 2020BRP57Z) "GEMS".

This research has made use of the VizieR catalogue access tool, CDS, Strasbourg, France (DOI : 10.26093/cds/vizier). The original description   of the VizieR service was published in 2000, A\&AS 143, 23"

\section*{Data Availability}

All raw data are publically available on the online repositories of the ESO, \textit{XMM-Newton} and \nustar{} archives.



\bibliographystyle{mnras}
\bibliography{example} 








\bsp	
\label{lastpage}
\end{document}